\documentstyle[twocolumn,prl,aps]{revtex}
\input epsf
\newlength{\myfigsize}
\begin{document}
\wideabs
{ 
\title
{
Short-Distance Correlation Properties 
of the Lieb-Liniger System
and 
Momentum Distributions of Trapped 
One-Dimensional Atomic Gases 
}
\author
{
Maxim Olshanii\cite{e-mail1}
and 
Vanja Dunjko  
}
\address
{
Department of Physics \& Astronomy,
University of Southern California,
Los Angeles, CA 90089-0484, USA\\
and\\
Institute for Theoretical Atomic and Molecular Physics,
Harvard-Smithsonian Center for Astrophysics,
Cambridge, MA 02138, USA
}
\date{\today}

\maketitle
\begin{abstract}
We derive exact 
closed-form expressions for the  
first few terms of the
short-distance Taylor expansion  
of the one-body correlation function of the
Lieb-Liniger gas.  
As an intermediate result
we obtain the high-$p$ asymptotics
of the momentum distribution of both
free and harmonically trapped atoms and show
that it obeys a universal $1/p^4$ law for {\it all} values of the 
interaction strength.
We discuss the ways to observe the predicted
momentum distributions experimentally,
regarding them as a sensitive identifier
for the Tonks-Girardeau regime of strong correlations.
\end{abstract}
}

%%%%%%%%%%%%%%%%%%%%%%%%%%%%%%%%%%%%%%%%%%%%%%%%%%%%%%%%%%%%%%%%%%%%%%%%%
%
{\it Introduction.}
Even though the correlation functions for 
the Lieb-Liniger gas of $\delta$-interacting 
one-dimensional bosons \cite{LL}
have been 
an object
of intense research in the Integrable Systems community
since the late
70s  \cite{Korepin},
the full closed-form expressions are known only in
the Tonks-Girardeau limit of infinitely strong interactions 
\cite{Vaidya}. 
While the scaling properties of the long-range \cite{short-long} asymptotics
of the correlation functions 
can be derived from Haldane's theory of quantum liquids
\cite{Haldane}, Conformal Field Theory \cite{conformal}, and 
Quantum Inverse Scattering method \cite{Korepin,Korepin_finite_g}, 
{\it virtually nothing} is known about short-range one-body correlations
at finite coupling strength \cite{Creamer-Jimbo}.
One of the goals of this paper is to extend
the existing knowledge in this direction. 

It is known that while for weak interactions 
the Lieb-Liniger system is well-described 
by the mean-field theory, the opposite, Tonks-Girardeau 
regime of infinitely strong interactions 
\cite{Tonks,Imp_Bosons} constitutes  
a strongly correlated system dual to a free Fermi gas. 
In experiments with one-dimensional atomic gases
\cite{Mott,1D_experiments} the one-body momentum distribution 
of the gas, along with the density profiles \cite{Vanja_Tonks}
and phase fluctuations \cite{Gora_phase_fluctuations,Gangardt,Stoof}, 
can readily help to distinguish 
between the two quantum regimes.  
In the Tonks-Girardeau limit, 
the momentum distribution for both free and harmonically confined gases was investigated by 
several authors \cite{Vaidya,Tonks_PRL,Marvin_big_Tonks,Anna_tail}. 
In this paper, we address the question of the momentum distribution 
in the intermediate, in between mean-field and Tonks-Girardeau, regime, as 
more realistic from the experimental point of view. 

%%%%%%%%%%%%%%%%%%%%%%%%%%%%%%%%%%%%%%%%%%%%%%%%%%%%%%%%%%%%%%%%%%%%%%%%%
%
{\it System of interest.}
Consider a one-dimensional gas of $N$ $\delta$-interacting bosons
confined in a length $L$ box with periodic boundary conditions. The Hamiltonian of 
the system reads
\begin{eqnarray}
\widehat{H}&=&
-\frac{\hbar^{2}}{2m}\sum_{j=1}^{N}\,\frac{\partial^{2}}{\partial z_{j}^{2}}
+
g_{\rm 1D}\sum_{i=1}^{N} \, \sum_{j=i+1}^{N} \, \delta \left(z_{i}-z_{j}\right)
\label{Hamiltonian_FQ}
\\
&=& \int_{-L/2}^{+L/2} \, \frac{\hbar^2}{2m}
\partial_{z}\hat{\Psi}^{\dagger}\partial_{z}\hat{\Psi} \, + \,
\frac{g_{\rm 1D}}{2} \hat{\Psi}^{\dagger}\hat{\Psi}^{\dagger}\hat{\Psi}\hat{\Psi}
\,,
\label{Hamiltonian_SQ}
\end{eqnarray}
where $m$ is the atomic mass, and $g_{\rm 1D}$ is the one-dimensional coupling constant,
whose expression for real atomic traps is given in \cite{Tonks_PRL}. 
This Hamiltonian can be diagonalized via Bethe ansatz \cite{LL}.
At zero temperature, the energy of the system    
is given through 
\begin{eqnarray}
E/N = \frac{\hbar^{2}}{2m}n^{2} e(\gamma)
\label{epsilon} \,\,\,,
\end{eqnarray}  
where the dimensionless parameter
$
\gamma = 2/n|a_{\rm 1D}| 
$
is inversely proportional to the one-dimensional
gas parameter $n \left|a_{\rm 1D}\right|$, $n$ is
the one-dimensional number density of particles,
$a_{\rm 1D} = -2\hbar^{2}/m g_{\rm 1D}$ is the 
one-dimensional scattering length 
introduced in \cite{Tonks_PRL},
and the function $e(\gamma)$ is
given by the solution of Lieb-Liniger system of equations \cite{LL}:
it is tabulated in \cite{table}.
Note the asymptotic behavior of $e(\gamma)$ (first computed in \cite{LL}):
\begin{eqnarray}
e(\gamma) \stackrel{\gamma \to 0}{\approx} \gamma  \hspace{3mm};\hspace{3mm}  
e(\gamma) \stackrel{\gamma \to \infty}{\approx} \frac{1}{3}\pi^2 \left(\frac{\gamma}{\gamma+2}\right)^2
\,,
\label{e(gamma)_limits}
\end{eqnarray}
where $\gamma \to 0$ corresponds to the mean-field or Thomas-Fermi regime, whereas 
$\gamma \to \infty$ corresponds to the Tonks-Girardeau regime.

%%%%%%%%%%%%%%%%%%%%%%%%%%%%%%%%%%%%%%%%%%%%%%%%%%%%%%%%%%%%%%%%%%%%%%%%%
%
{\it High-$p$ momentum distribution.}
Our first object of interest is the high-$p$ asymptotics of the one-body momentum 
distribution in the ground state.
To evaluate it, 
we need two mathematical facts, (a) and (b):

(a) The presence of the delta-function interactions in the Hamiltonian 
(\ref{Hamiltonian_FQ}) implies 
that its eighenfunctions undergo, at the point of contact of any
two particles $i$ and $j$, a kink in
the derivative proportional to the value of the eigenfunction 
at this point \cite{kink}:
\newpage
\begin{eqnarray} 
&&\Psi(z_1,\,\ldots,\,z_{i},\,\ldots,\,z_{j},\,\ldots,\,z_{N}) 
\nonumber
\\
&&\hspace{3mm} 
= \Psi(z_1,\,\ldots,\,Z_{ji},\,\ldots,\,Z_{ji},\,\ldots,\,z_{N})
\label{eq:contact}
\\
&&\hspace{6mm}
\times
\left\{
 1 - |z_{ji}|/a_{\rm 1D} + \varepsilon(|z_{ji}|;\, \{{\cal Z}_{ji}\})
\right\}
\nonumber
\\
&&\varepsilon(|z_{ji}|;\, \{{\cal Z}_{ji}\}) = {\cal O}(|z_{ji}|^2)
\,\,,
\nonumber
\end{eqnarray}   
where $Z_{ji} = (z_{i}+z_{j})/2$ and $z_{ji} = z_{j} - z_{i}$ are the center-of-mass and relative
coordinates of the $ij$ pair of particles, respectively, and
$\{{\cal Z}_{ji}\} = \{Z_{ji},\,z_1,\,\ldots,\,z_{i-1},\,z_{i+1},\,\ldots,\,z_{j-1},\,z_{j+1},\,\ldots,\,z_{N}\}$ 
denotes a set consisting of the center-of-mass coordinate of the $i$-th and $j$-th particles 
and the coordinates of all the other particles.

(b) Imagine that a periodic function $f(z)$, defined on the interval $[-L/2,\,+L/2]$, 
has a singularity of the form $f(z) = |z-z_{0}|^{\alpha}F(z)$, where $F(z)$ is a regular function,
$\alpha > -1$ and $\alpha \neq 0,\,2,\,4\,\ldots$.
Then the leading term in the asymptotics of the Fourier transform of $f$ reads
\cite{AsymptoticIntegral}
\begin{eqnarray}
&&\int_{-L/2}^{+L/2}dz \, e^{-ikz} \, f(z)
\nonumber
\\
&&\hspace{3mm}
\stackrel{|k|\to\infty}{=}
2\cos(\frac{\pi}{2}(\alpha+1))\, \Gamma(\alpha+1)\, e^{-ikz_{0}}\, F(z_{0})\, \frac{1}{|k|^{\alpha+1}}
\label{eq:AsymptoticIntegral}
\\
&&\hspace{6mm}
+ {\cal O}(\frac{1}{|k|^{\alpha+2}})
\,\,,
\nonumber
\end{eqnarray}
where $k = (2\pi/L)\, s$ and $s$ is an integer.
For multiple singular points of the same order, the full asymptotics is the sum of the corresponding 
partial asymptotics of the form (\ref {eq:AsymptoticIntegral}).

Let us evaluate, using  (\ref{eq:contact}) and (\ref{eq:AsymptoticIntegral}),
the momentum representation of the ground state wavefunction of the 
Hamiltonian (\ref{Hamiltonian_FQ}) with respect to the first particle:
\begin{eqnarray} 
&&\Psi(p_{1},\,z_{2},\,\ldots,\,z_{N}) 
\nonumber
\\
&&\hspace{2mm}
= L^{-\frac{1}{2}} \int_{-L/2}^{+L/2} dz_{1} e^{-p_{1}z_{1}/\hbar}
\Psi(z_{1},\,z_{2},\,\ldots,\,z_{N})
\nonumber
\\
&&\hspace{2mm}
\stackrel{\forall i:\,\, \, 2 \geq i \geq N}{=}
L^{-\frac{1}{2}} \int_{-L/2}^{+L/2} dz_{1} e^{-i p_{1}z_{1}/\hbar}
\nonumber
\\
&&\hspace{5mm}
\times 
\Psi(z_{1} = Z_{1i},\,\ldots,\, z_{i} = Z_{1i},\,\ldots,\,z_{N})
\nonumber
\\
&&\hspace{5mm}
\times
\left\{  
 1 - |z_{1i}|/a_{\rm 1D} + \ldots 
\right\}
\nonumber
\\
&&\hspace{2mm}
\stackrel{|p_{1}| \to \infty}{=}
\sum_{i=2}^{N} (2 L^{-\frac{1}{2}}/a_{\rm 1D}) e^{-i p_{1}z_{i}/\hbar}
\label{momentum_representation}
\\
&&\hspace{5mm}
\times
\Psi(z_{1} = z_{i},\,\ldots,\, z_{i},\,\ldots,\,z_{N})
\frac{1}{(p_{1}/\hbar)^2}
\nonumber
\end{eqnarray}
Here $p_{1} = (2\pi\hbar/L)\, s$, where $s$ is an integer.

Let us now turn to the one-body momentum distribution 
{\it per se}. After a lengthy but straightforward
\cite{3-body_terms}
calculation it takes the form
\begin{eqnarray}
w(p) 
&\equiv& 
\int_{-L/2}^{+L/2} dz_{2}\,\ldots\, \int_{-L/2}^{+L/2} dz_{N}
|\Psi(p,\,z_{2},\,\ldots,\,z_{N})|^2
\nonumber
\\
&\stackrel{|p| \to \infty}{=}&
\frac{4(N-1)\rho_{2}(0,0,0,0)}{a_{\rm 1D}^2}\,\frac{1}{(p/\hbar)^4}
\,\,,
\label{momentum_distribution_intermediate}
\end{eqnarray}
where $\rho_{2}(z_{1},\,z_{2};\,z_{1}^{\prime},\,z_{2}^{\prime})$ is the 
two-body density matrix, normalized as 
$
\int_{-L/2}^{+L/2} dz_{1} \, \int_{-L/2}^{+L/2} dz_{2} \,
\rho_{2}(z_{1},\,z_{2};\,z_{1},\,z_{2}) = 1
$,
and 
$w(p)$ is the momentum distribution, 
normalized as 
$\sum_{s=-\infty}^{+\infty} w(2\pi \hbar s / L) = 1$.

The expression (\ref{momentum_distribution_intermediate}) involves the two-body 
density matrix whose form is unknown for a finite system. However, 
an elegant thermodynamic limit formula for $\rho_{2}(0,0,0,0)$ does exist
due to Gangardt and Shlyapnikov \cite{Gangardt}, who derived it using the 
Hellmann-Feynman theorem \cite{Hellmann-Feynman}: $L^2\, \rho_{2}(0,0,0,0) = e^{\prime}(\gamma)$. 
We are now ready to give a closed-form thermodynamic limit expression for the 
high-$p$ asymptotics of the one-body momentum distribution for one-dimensional $\delta$-interacting bosons in a 
box with periodic boundary conditions:
\begin{eqnarray}
W(p) 
\stackrel{|p| \to \infty}{=}
\frac{1}{\hbar n} \, \frac{\gamma^2  e^{\prime}(\gamma)}{2\pi} \, \left(\frac{\hbar n}{p}\right)^4
\,,
\label{momentum_distribution}
\end{eqnarray}
where $W(p) = (L/2\pi\hbar)\, w(p)$ is normalized as $\int_{-\infty}^{+\infty} dp \, W(p) = 1$.
Notice that this asymptotics is {\it universally} described by a $1/p^4$ law for all values 
of the coupling strength $\gamma$. (Note that for $\gamma\to\infty$, this law was predicted 
in \cite{Anna_tail}.)  
Formula (\ref{momentum_distribution}) is {\it the first  of the 
two principal results of our paper}.

\begin{figure}
\setlength{\myfigsize}{0.47\textwidth}
\epsfxsize=\myfigsize
\begin{center}
\parbox{\myfigsize}{\epsfbox[68 48 384 290]{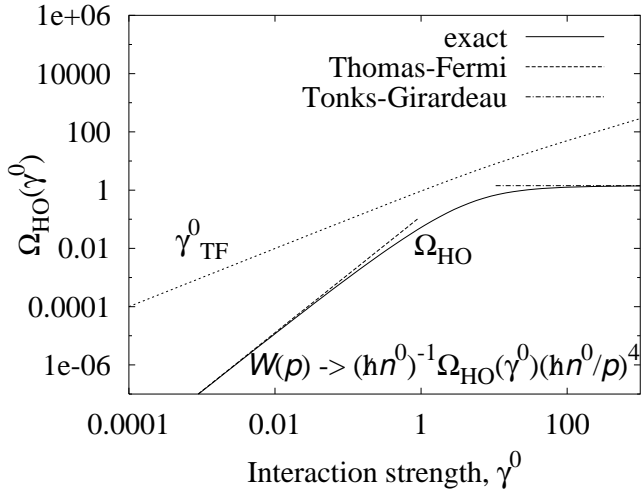}}
\end{center}
\caption
{
Dimensionless coefficient $\Omega_{HO}(\gamma^{0})$
in front of the leading term  of the
high-$p$ asymptotics of 
the momentum distribution of {\it harmonically} trapped atoms, as a function of the
interaction strength $\gamma^{0}$ in the center of the cloud. Also shown
is the directly related to the experimental parameters
Thomas-Fermi estimate $\gamma^{0}_{\rm TF} = (8/3^{2/3}) (N m a_{\rm 1D}^2 \omega / \hbar)^{-2/3}$
for the interaction strength in the center, as a function of $\gamma^{0}$.
Both momentum and momentum distribution are measured in units related
to the density in the center $n^{0}$, that can be expressed through
$\gamma^{0}$ using $\gamma^{0} = 2/n^{0} a_{\rm 1D}$.
}
\label{fig:HOLeadMomCoef}
\end{figure}

%%%%%%%%%%%%%%%%%%%%%%%%%%%%%%%%%%%%%%%%%%%%%%%%%%%%%%%%%%%%%%%%%%%%%%%%%
{\it Harmonically trapped 1D gas: momentum distribution.}
To evaluate the high-$p$ asymptotics of the 
momentum distribution of atoms confined in a harmonic 
oscillator potential, we are going to employ the 
local density approximation (LDA). It is based on an intuitive,
but hard to justify rigorously assumption that in the thermodynamic limit the short-range 
correlation properties of a trapped gas are indistinct from the 
ones of a uniform gas of the same local density:
$
\sigma(z,\,z^{\prime}) \stackrel{z-z^{\prime} \ll l}{\approx}
\sigma_{u}(z-z^{\prime} \, | \, n((z+z^{\prime})/2)
$, 
where $\sigma(z,\,z^{\prime})$ and 
$\sigma_{u}(z-z^{\prime} \, | \, n)$ are  
one-body density matrix of the trapped gas and one-body 
density matrix of a density $n$
uniform gas 
respectively, both normalized to the respective number of particles, 
$n(z)$ is the density profile of the 
trapped gas, $l$ is the typical length on which the density 
changes. 
(Tested against the exact results on correlation function of 
the trapped Tonks-Giradeau gas \cite{Papenbrock} this assumption
can be shown to lead to an exact prediction for the value 
of the coefficient in front of the $(z-z^{\prime})^2$ term in the 
Taylor expansion around $z^{\prime} = -z \approx 0$.)
From this ansatz it immediately follows that 
the high-$p$ asymptotics of the momentum distribution 
(sensitive to the short-range correlations only) 
is given by the {\it spatial average} 
of the uniform case expression (\ref{momentum_distribution}) over the  density 
profile 
of the atomic cloud. 
The density profiles themselves can also be obtained using LDA
(\cite{Vanja_Tonks}, Eqns. 21 and 22, where the governing parameter
$\eta$ should be replaced by $2/\gamma^{0}_{\rm TF}$, see below), and this is the method we used.
The final result is presented in Fig.\ref{fig:HOLeadMomCoef}.
There the dimensionless coefficient $\Omega = \lim_{|p|\to\infty} W(p) p^4 /(\hbar n^{0})^3$  
in front of the high-$p$ asymptotics of the momentum distribution
is plotted as a function of the interaction strength parameter
$\gamma^0$ in the center of the cloud; $\gamma^{0}$ in turn depends on  
of the experimental parameters, such as the number of particles $N$, the coupling constant $g_{\rm 1D}$, 
and the longitudinal trap frequency $\omega$, through a system of implicit  
equations \cite{Vanja_Tonks}. Here $n^{0}$ is the density in the center of the trap.
To establish a link to the experimental parameters
we also present a plot for the Thomas-Fermi (weak interactions) prediction 
for $\gamma$ in the center of the atomic cloud, $\gamma^{0}_{\rm TF} = (8/3^{2/3}) (N m a^2 \omega / \hbar)^{-2/3}$, 
as a function of $\gamma^{0}$.

In the limiting, Thomas-Fermi and Tonks-Girardeau regimes the momentum distribution is given by
\begin{eqnarray}
&&
\displaystyle
W(p) \stackrel{|p|\to\infty,\,\,\gamma \to 0}{\approx} 
\frac{1}{p_{\rm HO}}
\frac{2\cdot 3^{\frac{2}{3}}}{5\pi} N^{\frac{2}{3}}
\left(\frac{a_{\rm HO}}{|a_{\rm 1D}|}\right)^{\frac{5}{2}} \left(\frac{p_{\rm HO}}{p}\right)^4 
\nonumber
\\
&&
\displaystyle
W(p) \stackrel{|p|\to\infty,\,\,\gamma \to \infty}{\approx} 
\frac{1}{p_{\rm HO}}
\frac{\sqrt{2}\cdot 128}{45 \pi^3} N^{\frac{3}{2}} \left(\frac{p_{\rm HO}}{p}\right)^4
\label{W(p)_limits}
\end{eqnarray}
where $a_{\rm HO} = (\hbar/m\omega)^{1/2}$ and $p_{\rm HO} = \hbar/a_{\rm HO}$.
%---------------------------------------
%---------------------------------------
%

%%%%%%%%%%%%%%%%%%%%%%%%%%%%%%%%%%%%%%%%%%%%%%%%%%%%%%%%%%%%%%%%%%%%%%%%%
%
{\it Short-range expansion for the correlation function.}
Let us now redirect our attention to the ground state one-body correlation function
\begin{eqnarray}
g_{1}(z) = \langle \hat{\Psi}^{\dagger}(z) \hat{\Psi}(0) \rangle \,,
\end{eqnarray}  
and in particular to its Taylor expansion around zero:
\begin{eqnarray}
g_{1}(z)/n = 1 \, + \, \sum_{i=1}^{\infty} c_{i} |nz|^i 
\,.
\label{correlation_function}
\end{eqnarray}
In the limit of infinitely strong interactions $\gamma \to \infty$, 
this expansion is known to all orders \cite{Vaidya}:
\begin{eqnarray}
c^{{\rm TG}}_{1} = 0 \,;\,
c^{{\rm TG}}_{2} = -\frac{\pi^2}{6} \,;\,
c^{{\rm TG}}_{3} = \frac{\pi^2}{9} \,;\,
c^{{\rm TG}}_{4} = \frac{\pi^4}{120} \,;\,
\ldots 
\label{c_1-4_Tonks}
\end{eqnarray}
Our goal now is to obtain the first few (through the order $|z|^3$) coefficients 
of the expansion (\ref{correlation_function})
for an {\it arbitrary} interaction strength $\gamma$.

The knowledge of the momentum distribution
(\ref{momentum_distribution}) is crucial for determining the 
$c_{1}$ and $c_{3}$ coefficients.
Let us look at the relation between the momentum distribution and the correlation function,
where the former is simply the Fourier transform of the latter:
$
W(p) = (2\pi\hbar n)^{-1} \int_{-\infty}^{+\infty}\!dz \, e^{-ipz/\hbar} g_{1}(z)
$.
Since the leading term in the asymptotics of $W(p)$ is $1/p^4$ we may conclude, 
using the Fourier analysis theorem (\ref{eq:AsymptoticIntegral}), that the lowest
{\it odd} power in the short-range expansion of the correlation function $g_{1}(z)$ is
$|z|^3$, and therefore the $|z|$ term is absent from the expansion: 
\begin{eqnarray}
c_{1} = 0
\,\,.
\label{c_1}
\end{eqnarray}
Furthermore, the theorem (\ref{eq:AsymptoticIntegral})
allows one to deduce the coefficient $c_{3}$ from the momentum
distribution (\ref{momentum_distribution}): 
\begin{eqnarray}
c_{3} = \frac{1}{12}\, \gamma^2 \, e^{\prime}(\gamma)
\,\,.
\label{c_3}
\end{eqnarray}

To obtain the coefficient $c_{2}$, we employ
the Hellmann-Feynman theorem \cite{Hellmann-Feynman} again.
Let a Hamiltonian $\hat{H}(w)$ depend on a parameter $w$. Let
$E(w)$ be an eigenvalue of this Hamiltonian. Then the mean value
of the derivative of the Hamiltonian with respect to the parameter
can be expressed through the derivative of the eigenvalue:
$
\langle \Psi_{E}(w) | \frac{d}{dw}\,\hat{H}(w) | \Psi_{E}(w) \rangle
=
\frac{d}{dw} E(w)
$. Let us now denote the fraction $\hbar^2/m$ as $\kappa$ and
differentiate the Hamiltonian (\ref{Hamiltonian_SQ}) with respect 
to $\kappa$. According to the Hellmann-Feynman theorem, we 
get 
$
\frac{1}{2}
\int_{-L/2}^{+L/2} dz \,
(\partial^2/\partial z \partial z^{\prime}) 
\langle 
  \hat{\Psi}^{\dagger}(z)\, \hat{\Psi}(z^{\prime}) 
\rangle
\Big|_{z^{\prime}=z}
= dE / d\kappa 
$.
Now, using 
$\langle \hat{\Psi}^{\dagger}(z)\, \hat{\Psi}(z^{\prime}) \rangle = 
\langle \hat{\Psi}^{\dagger}(z-z^{\prime})\, \hat{\Psi}(0) \rangle$, we obtain
$- \frac{1}{2} \, L \, [(d^2/dz^2) g_{1}(z)]\Big|_{z=0}   = d E/ d \kappa$, and finally 
\begin{eqnarray}
c_{2} = - \, \frac{1}{2} \, 
\left\{
e(\gamma) \, - \, \gamma\,e^{\prime}(\gamma)
\right\}
\,,
\label{c_2}
\end{eqnarray}
where we have used the {\it known} expression for the energy (\ref{epsilon}).

Note that, as expected, our expressions for the coefficients $c_{1-3}$  
converge, in the limit $\gamma \to \infty$, to the known 
results for the impenetrable bosons (\ref{c_1-4_Tonks}). 
This can be easily verified using the 
$\gamma \to \infty$ expansion for the function $e(\gamma)$
(\ref{e(gamma)_limits}).

Expressions ({\ref{c_1}), (\ref{c_2}), and (\ref{c_3}) constitute the
{\it second principal result obtained in our paper}. 

%%%%%%%%%%%%%%%%%%%%%%%%%%%%%%%%%%%%%%%%%%%%%%%%%%%%%%%%%%%%%%%%%%%%%%%%%
%
{\it Concluding remarks.} Below we present a discussion 
on empirical observation of and applications for 
the $1/p^4$ momentum distribution tails, 
in experiments with {\it harmonically trapped}
atomic gases.
 
(a) First of all, we would like to discuss   
the momentum range where the $1/p^4$ tail should be looked for 
experimentally. Relying on the $\gamma \to \infty$ results 
\cite{Korepin} and an analysis of the predicted-by-Bogoliubov's-theory momentum 
distribution (corresponding to $\gamma \to 0$), we conjecture that in
the whole range of the interaction strength $\gamma$, the high-$p$
asymptotics of the momentum distribution corresponds to the range of momenta given by
$p \gg (m\mu/\hbar^2)^{1/2}$ for all $\gamma$, where $\mu$ is the chemical potential
of the system. For the case of a harmonically confined
gas at $\gamma \begin{array}{c}>\vspace{-2mm}\\\sim\end{array} 1$ this leads to 
\begin{eqnarray} 
p \gg \sqrt{N} p_{\rm HO}&&\hspace{3mm}\mbox{for }\gamma\begin{array}{c}>\vspace{-2mm}\\\sim\end{array} 1
\,.
\label{range}
\end{eqnarray}

(b) Our zero-temperature results are valid as long as the temperature does 
not exceed the chemical potential, $k_{\rm B}T \ll \mu$ (for all $\gamma$), or, for the Tonks-Giradeau case, 
\begin{eqnarray}  
k_{\rm B}T \ll N \hbar\omega  
&&\hspace{3mm}\mbox{for }\gamma\begin{array}{c}>\vspace{-2mm}\\\sim\end{array} 1
\,.
\end{eqnarray}   
For temperatures comparable to the chemical potential, the 
$1/p^4$ law should persist within the range (\ref{range}), 
but the prefactor is not yet known and 
it is a subject of future research.

(c) Experimentally, the momentum distribution of the Tonks-Girardeau gas 
can be detected either 
{\it in situ} \cite{Marvin_Raman} or via a ballistic expansion \cite{Markus_private}. The latter 
option requires some caution, especially in the Tonks-Giradeau ($\gamma\to\infty$) case. 
If only the longitudinal confinement is released, the interactions will modify the 
momentum distribution during the expansion, and, as follows from the Bose-Fermi 
mapping \cite{Imp_Bosons}, the detected momentum
distribution will correspond to a free Fermi gas. Instead one should release 
both longitudinal and transverse confinements simultaneously. In this case 
rapid transverse ballistic expansion will lead to a quick drop-off of the density, 
making the interactions negligible.

(d) We believe that the coefficient in front of the 
high-$p$ tail of the momentum distribution 
(Fig.\ref{fig:HOLeadMomCoef})
may provide a {\it robust} experimental identifier of the quantum regime 
of the gas of interest, and, in particular, serve to detect the 
Tonks-Girardeau regime. (i) The high-$p$ tail is not sensitive 
to the finite temperature corrections to the correlation function,
which appear predominantly in the low-$p$ (long-range) domain. 
(ii) In experiments with 2D optical lattices, where a single 
cigar-shaped trap is replaced by an array of traps \cite{Mott},
the effect of the residual 3D mean-field pressure acting during the expansion 
becomes relevant: the high-$p$ part of the momentum distribution is 
far less sensitive to this effect as compared to the low-$p$ part.
(iii) The theoretical interpretation of the experimental results is 
simpler in the high-$p$ case thanks to the applicability of 
the LDA. In the opposite low-$p$ case, the LDA leads to entirely 
wrong predictions \cite{Chiara_private}. 

%%%%%%%%%%%%%%%%%%%%%%%%%%%%%%%%%%%%%%%%%%%%%%%%%%%%%%%%%%%%%%%%%%%%%%%%%
%
{\it Summary.} In this paper, we present a short-range Taylor expansion  
(up to the order $|z|^3$)
for the zero-temperature correlation function $g_{1}(z)$ of a one-dimensional
$\delta$-interacting Bose gas (see Eqns.\ \ref{correlation_function}, \ref{c_1}, \ref{c_2}, and \ref{c_3}). 
We compute the leading term in the high-$p$ asymptotics of the momentum distribution
for both free (Eqn.\ \ref{momentum_distribution}) and 
harmonically trapped (Fig.\ref{fig:HOLeadMomCoef}) atoms. We regard the high-$p$ tail of the momentum 
distribution as an efficient tool for identification of the Tonks-Girardeau regime
in experiments with dilute trapped atomic gases.

%%%%%%%%%%%%%%%%%%%%%%%%%%%%%%%%%%%%%%%%%%%%%%%%%%%%%%%%%%%%%%%%%%%%%%%%%

{\bf Acknowledgments}. Authors are grateful to 
M. Girardeau, D.M. Gangardt and C. Menotti
for enlightening discussions on the subject. 
M.\ O.\ would like to acknowledge the hospitality 
of the European Centre for Theoretical Studies in Nuclear 
Physics and Related areas (ECT$\star$) during the 2002 BEC Summer Program, 
where the presented work was initiated. 
This work was supported by 
the NSF grant {\it PHY-0070333} and ONR grant {\it N000140310427}. 
Authors appreciate financial support by NSF through the 
grant for Institute 
for Theoretical Atomic and Molecular Physics, Harvard Smithsonian Center 
for Astrophysics.

%%%%%%%%%%%%%%%%%%%%%%%%%%%%%%%%%%%%%%%%%%%%%%%%%%%%%%%%%%%%%%%%%%%%%

\begin{thebibliography}{99}
\bibitem{short-long}
The boarder between short and long range is given by 
the correlation length $l_{c} \sim \hbar/\sqrt{m \mu}$, $\mu$ being the 
chemical potential.
\bibitem{3-body_terms}
The double sum over particles involved in the computation
can be split into a diagonal part (proportional to the two-body
density matrix) and an off-diagonal part involving Fourier
transforms of the three-body density matrix. The latter 
can be shown to decay faster than $1/p^4$ and thus does not 
contribute to the final result.
\end{thebibliography}
\end{document}